\documentclass[12pt]{article}
\usepackage{english}
\usepackage{epsfig}
 \newcommand{\beq}{\begin{equation}}
 \newcommand{\eeq}{\end{equation}}
 \newcommand{\beqa}{\begin{eqnarray}}
 \newcommand{\eeqa}{\end{eqnarray}}
 
 \def\pabl#1#2{\frac{\partial {#1}}{\partial { #2}}}

 \def\dabl#1#2{\frac{{\rm d}{#1}}{{\rm d}{#2}}}
 
\begin{document}

\begin{center}

%
%
{\Large{\bf Stochastic relaxational dynamics applied to finance: towards non-equilibrium option pricing theory }}\\[2cm]

{\large{ Matthias Otto}}\\
 Institut f\"ur Theoretische Physik, 

Universit\"at G\"ottingen, Bunsenstrasse 9, D-37073 G\"ottingen,
Germany,\\
e-mail: matthias.otto@physik.uni-goettingen.de  \\ [2cm]


\underbar{Abstract}

\end{center}
Non-equilibrium phenomena occur not only in physical world, but
also in finance. In this work, stochastic
relaxational dynamics (together with path integrals) is applied to option
pricing theory. Equilibrium in financial markets is
defined as the absence of arbitrage, i.e. profits ``for nothing''.
A recently proposed model (by Ilinski et al.) considers fluctuations
around this equilibrium state by introducing
a relaxational
dynamics with random noise for intermediate 
deviations called 
``virtual'' arbitrage returns. 
In this work, the model is
incorporated within a martingale
pricing method for derivatives on securities (e.g. stocks) in
incomplete markets 
using a mapping to option
pricing theory with stochastic interest
rates. 
The arbitrage return is considered as a component of a fictitious short-term interest
rate 
in a virtual world. The influence of intermediate
arbitrage returns on the price of 
derivatives in the real world can be recovered by performing an
average over the (non-observable) arbitrage return at the time of
pricing. 
Using a famous result by Merton and with some help
from the path integral method,
exact pricing formulas for European call and put options under the
influence of virtual arbitrage returns (or intermediate deviations
from economic equilibrium) are derived where only the final integration
over initial arbitrage returns needs to be performed numerically. This result, which has not been
given previously and is at variance with results stated by Ilinski et al., 
is complemented by a discussion of the hedging strategy associated to
a derivative, which replicates the final payoff but turns out to
be not self-financing in the real world, but self-financing {\it when
  summed over the derivative's remaining life time}. 
Numerical examples are given which underline the fact that 
an additional positive risk premium (with respect to the
Black-Scholes values) is found reflecting extra hedging costs due to
intermediate deviations from economic equilibrium.

\newpage
\section{Introduction}
The pricing and hedging of derivatives is a major task for financial institutions \cite{hull:97} and has become 
an increasingly popular topic in statistical physics \cite{bouchaud:94,bouchaud:98}.
Derivatives are sometimes also called contingent claims, as the buyer
of the derivatives is entitled to receive a certain payoff up to (or
at) some future time $T$, the time of expiry, dependant on the price
$S$ of a so-called ``underlying'' security (say a stock) within a
certain time interval between today and $T$ or at time $T$. 
The simplest case is a so-called European call (put) option which gives the buyer the right
to buy 
(sell) an underlying security (e.g. a stock) at a certain
time $T$ in the future for a fixed price $K$ (the strike
price). These options are also called ``plain vanilla options'' for
their simplicity (as common as the vanilla icecream flavor).

The classical result of Black and Scholes \cite{black:73} on option pricing
which revolutionized the world of finance and still forms the
foundation for most of modern research, is based on the existence of
an equilibrium, generally called ``absence of arbitrage'', i.e. the
impossibility of a profit ``for nothing''. 

The use of the
no-arbitrage assumption for pricing purposes is nicely elucidated in a standard
text book like \cite{hull:97} where simple pricing equations for forward
contracts are derived from optimization arguments. If e.g. the forward price $F$ at time $t$
for buying or selling (assuming no bid/offer spreads and transactions
costs) a non-dividend
paying security $S$ at a later time $T$ were less than
$S\exp(r(T-t))$, then a riskless profit could be
obtained in the following way: at time $t$, one enters into a forward contract to buy the
security for $F$ at time $T$, and one short sells the security
(i.e. one borrows the security from somebody else and sells it,
assuming no fees for simplicity) and puts the
proceeds on a deposit at the riskless interest rate $r$ (assuming no
credit risk); at time $T$, one receives the security from the forward
contract thereby closing out the short position  in the security
(i.e. handing it over to the lender), while
receiving the nominal amount plus interest from the deposit less the
forward price paid: $S\exp(r(T-t))-F>0$. Likewise, an arbitrage is
possible when $F>S\exp(r(T-t))$. As the information on either
situation spreads in the market, the inequalities disappear, and the
relation $F=S\exp(r(T-t))$ results. Now the no-arbitrage assumption
anticipates the equality to hold right from the beginning, thus
implying that arbitrage opportunities disappear infinitely fast.  
Now many trading activities are motivated exactly by the fact, that
this is not the case, but that arbitrage returns exist in
the market for a short time $\tau_{\rm arbitrage}>0$. After this time,
the information on arbitrage opportunities has reached enough market
participants to make them disappear. 

The absence of arbitrage assumption paves the way for one of the
 fundamental pillars of  mathematical finance which  is
the theorem by Harrison and Pliska \cite{harrison:81}. In fact whenever markets are
complete (i.e. when any derivative can be hedged by a self-financing
strategy, which is a more restrictive statement than absence of arbitrage),
then there is a unique
equivalent martingale measure for the underlying security and vice
versa (see \cite{baxter:96} for an introductory discussion on
martingale theory). A stochastic process $X_t$ is a martingale with
respect to the measure ${\rm Q}$ if and only if $E_{\rm Q}\left[|X_t|\right]<\infty$
\beq
X_t=E_{\rm Q}\left[X_s|{\cal F}_t\right],\qquad s\geq t
\eeq
where ${\cal F}_t$ is the filtration at time $t$, i.e. the information accumulated until time $t$.
This rather
technical statement is the basis for risk-neutral valuation:
Derivatives can be priced in a world where all yields are equal to the
risk-free interest rate (minus dividend yields etc.). 

Apart from the dynamic deviations from the no-arbitrage situation
discussed above, there is a large literature on
 serious drawbacks of the Black-Scholes model
itself which is classically used to implement no-arbitrage pricing of options,
i.e. that price returns evolve according to Brownian motion with constant drift and volatility. Empirical
studies of return distributions in fact show volatility clustering
and fat tails \cite{mandelbrot:63,mantegna:95,ghash:96,bouchaud:97}, and so real price changes appear to be more efficiently
modelled by truncated L\'evy processes (TLP) \cite{mantegna:95,koponen:95}. 
However, rational option pricing using the martingale approach appears
to be still working (see \cite{boyar:99} and ref.s therein), so the
main theme of the Black-Scholes method, i.e. the possibility to set up
a self-financing hedging strategy (a notion to be explained below) seems to hold.
Moreover, studies on
autocorrelations of price changes or on the distributions of price changes
themselves (taken for different time scales) demonstrate a crossover
to gaussian dynamics after a certain time scale which might vary from
several minutes to days (depending on market
liquidity)\cite{ghash:96,bouchaud:97}. 

In the present work, this time scale is proposed to be proportional to
$\tau_{\rm arbitrage}$. In this sense, fat tails of distributions of
price changes are a signature of intermediate arbitrage opportunities
on short time scales (compared to $\tau_{\rm arbitrage}$). The
implications for pricing options that are not very short-lived seem
to be that intermediate arbitrage opportunities may be modelled as
deviations from Brownian motion, and thus as perturbative effects. 
One way to treat these deviations are
stochastic volatility (SV) models (see for a review in the context of option
pricing \cite{frey:96}). The present work complements these models and
gives an alternative approach which
modifies the drift of the asset price process rather than its volatility.

If one tries to get rid of the drawbacks of the Black-Scholes model by
dynamic parameters which are not directly tradable such as in SV
models or in the approach discussed below, one encounters a new
problem: as opposed to complete markets defined above, a
self-financing strategy using traded instruments ceases to
exist. In general, any hedging strategy can only reduce the risk inherent
in the final payoff to an intrinsic component \cite{foelmer:91,bouchaud:94}. Technically, this leads to more than one equivalent martingale
measure \cite{harrison:81}.
ambiguity for derivatives pricing is handled by introducing
additional constraints on the hedging strategy, e.g. minimizing the
expected squared cost for the remaining life time of the option while
exactly replicating the final payoff (local risk-minimization)
or minimizing the expected squared net loss at the time of maturity of
the option (mean-variance hedging) \cite{foelmer:91,heath:98,fedotov:98} (see
also \cite{bouchaud:94,bouchaud:98} for a physicist's approach).
For the model presented in this work, a very specific method is
proposed in order to select an equivalent martingale measure which
satisfies both constraints simultaneously.

The issue of option pricing in incomplete markets has become a matter
of practical interest recently, in particular with the increasing importance of
credit derivatives. 
As opposed to conventional
derivatives which cover market risks, there is not a large underlying
market of actively traded credit risk instruments in every credit risk
category. 
As opposed to a
stock, e.g. a loan is usually not traded. 


Considering fully complete and incomplete markets as extreme cases of
real markets, one is naturally forced to ask for crossover effects or transitions
between the two regimes. A possible answer to this question might be
given in terms of a dynamic model which considers market incompleteness in
terms of fluctuations around an economic equilibrium characterizing a
complete market.

An important step in this direction has been given by
Ilinski and Stepanenko \cite{ilinski:99a} and
Ilinski \cite{ilinski:99b} who assume the existence of intermediate,
``virtual'' arbitrage returns $x_t$, which appear and disappear over a certain
time scale which may be identified with $\tau_{\rm arbitrage}$
mentioned above.
In fact, Ilinski et al. intend to treat arbitrage effects
as a perturbation to the usual Black-Scholes risk-free rate $r$, which
gives the
yield on all investments in a risk-neutral world. More specifically, the
Black-Scholes risk-free rate is split into a constant part $r^0$ and an
arbitrage return $x_t$ according to $r_t=r^0+x_t$. 
This model is different from stochastic volatility models: it
modifies the (risk-adjusted) drift of the asset price process rather than its
volatility. In principle there appears no reason to favor one approach
over the other. In fact, preliminary results from simulations of the
stochastic drift model versus the stochastic volatility model by Stein
\& Stein \cite{stein:91} seem to point in this direction. 
Detailed results on this comparison will be published elsewhere \cite{otto:99b}.
The stochastic drift approach discussed here has the
advantage that it can be mapped to models with stochastic interest
rates (see below).

Ilinski et al. \cite{ilinski:99a}\cite{ilinski:99b} present two
approaches to calculate option prices. In \cite{ilinski:99a}, a
perturbative method is given based on the classical Black-Scholes
equation where the constant risk-free rate $r$ is replaced by
$r^0+x_t$ where $r^0$ is constant and $x_t$ is random. This equation
is then iterated to second order in $x_t$ and averaged over with
respect to $x_t$. The results obtained for European call and put
options are not reproduced by our exact calculation. We come back to
this difference at the end of section 5 which explains why our results
are reasonable for the specific dynamics of the arbitrage return used
here. The observed difference just mentioned leads
us to suggest that either the method or some further approximations
that made in \cite{ilinski:99a} are not correct. In a second paper
\cite{ilinski:99b} Ilinski et al. derive a fully deterministic Black-Scholes type PDE that
depends both on the current level of the security price $S_t$ and the
arbitrage return $x_t$. This equation is not further evaluated. In
section 3, we show that it contains a misprint.
We add as a remark that the issue of measure change and the construction and
selection of an equivalent martingale measure which is fundamental
to option pricing (see e.g. \cite{baxter:96}) is not addressed. 

Therefore, in this article a different route is proposed. First, the
arbitrage return $x_t$
is considered as a part of a stochastic interest rate dynamics for the
risk-free rate $r_t$ in a virtual world (where arbitrage returns are
directly 
observable). 
Essentially, one
is led to a Black-Scholes type equation for a derivative depending on
two state variables, the security price $S_t$ and the arbitrage return
$x_t$ which
is derived from the risk-free rate in the virtual world according to 
$r_t=r^0+x_t$. The constant part $r^0$ is supposed to be the 
risk-free rate in the real
world which consequently is assumed to be constant. 
The reason for the latter simplification is the later comparison with the
Black-Scholes pricing formulas. 
The implementation of $x_t$ as a part of a fictitious interest rate
process leads to a stochastic drift for the asset process $S_t$ (with
respect to the particular martingale measure chosen, for details see
below) and thus couples the dynamics of $x_t$ to $S_t$.
As the arbitrage return is an intermediate phenomenon on time
scales shorter than the time to expiry, we follow
Ilinski et al. by enforcing the boundary condition at the time of
expiry of the option that the arbitrage return should
disappear. This constraint may be relaxed, however, if one allows for the possibility that any 
hedging strategy might not replicate (i.e. provide for) the final payoff of the option. 
Nonetheless, we stay with this constraint (also in order to compare
with the results of Ilinski et al. \cite{ilinski:99a}\cite{ilinski:99b}). However, we do not implement this condition into the payoff
function of the option like Ilinski \cite{ilinski:99b}, but into the
average over arbitrage returns. This procedure allows us to use a
famous result by Merton on options in a stochastic interest rate
environment \cite{merton:73}. 
 It is not meant to imply that intermediate arbitrage returns
can be thought of as the random part of real interest rates.
After averaging, we obtain a previously unknown
exact result for European claims under the influence of virtual
arbitrage. (The pricing of American claims which may be exercised
prior to their maturity is possible in principle using
a ``tree'' procedure \cite{hull:97}, i.e. a scheme based on discrete probabilities
and discrete time). These exact pricing formulas differ significantly from the
results obtained by Ilinski et al.\cite{ilinski:99a}. 
 
The outline of this paper is as follows:
In the next section, we present the route from the Black-Scholes model
to a non-equilibrium market model, taking up the idea of intermediate (``virtual'') arbitrage by Ilinski
and Stepanenko. The third section will show how the
effect of arbitrage returns on option pricing can be considered in
terms of a stochastic interest rate environment in a virtual
world. In section 4, European call and put options are valued in the
presence of virtual arbitrage returns. In section 5, the issue of a 
replicating hedging strategy both in the virtual and real world and
the selection of an equivalent martingale measure
is addressed. Some explicit numerical pricing examples are given and
their difference to the classical Black-Scholes results are explained in
section 6. In the final section, the results are briefly discussed.

\section{From the Black-Scholes model to the dynamics of arbitrage returns}
Let us briefly review the Black-Scholes analysis in order to motivate
the notion of arbitrage returns, following \cite{ilinski:99a}.
The model for a one security market is given by
\beq
\label{S}
dS_t=\mu S_tdt + \sigma S_t dW^1_t
\eeq
where $S_t$ is the security price, $\mu$ the drift, and $dW^1_t$ a
Wiener process. It may be motivated from the fact that $\ln
(S_{i+1}/S_i)$, where $i+1$ and $i$ denote discrete points in time,
performs a random walk \cite{hull:97}.  
Now 
the price of a derivative $V_t=V(S_t,t)$ whose payoff is contingent on the
security price $S_T$ at some future time $T$ can be
determined by setting up a portfolio $\Pi_t$ consisting of the derivative
$V_t$ and a position  $-\Delta$ of the security $S_t$:
\beq
\Pi_t=V_t-\Delta S_t
\eeq
Then if $\Delta=\pabl{V}{S}$, where $S=S_t$, this portfolio is riskless
as uncertainties arising from the Wiener process are eliminated which
can be seen by evaluating $d\Pi_t$ using Ito's lemma. Therefore, the
portfolio is known to grow at the risk-free rate, i.e.
\beq
\label{riskfree}
d\Pi_t=r\Pi_t dt
\eeq
For constant interest rates, equating expressions for $d\Pi_t$ gives the
Black-Scholes partial differential equation (PDE):
\beq
\label{BS}
\pabl{V}{t}+\frac{\sigma^2 S^2}{2}\pabl{^2 V}{S^2}+rS\pabl{V}{S}-rV=0
\eeq
Specifying a certain boundary condition to this equation representing
the option payoff at the time of maturity completes the usual
Black-Scholes pricing problem. As a reminder, let us note that the drift $\mu$ which
was introduced in the market model Eq.(\ref{S}) is absent from the
pricing equation Eq.(\ref{BS}).

The idea of arbitrage returns may be motivated by assuming that in the
presence of arbitrage opportunities, the true return of the portfolio
$\Pi_t$ is not equal to the constant risk-free interest rate $r$, but
might be less or more than that.
Following \cite{ilinski:99b}, an arbitrage return
$x_t$ is now introduced by substituting for $r$ 
\beq
\label{arb.0}
r_t=r^0+x_t
\eeq
where $x_t$ is assumed to follow the dynamics of a decay process with
random noise:
\beq
\label{arb}
\dabl{x_t}{t}=-\lambda x_t+\eta_t
\eeq
where $\eta_t$ is characterized by:
\beq
\langle\eta_t\rangle=0,\;\;\;\;\;\;\langle\eta_t\eta_{t'}\rangle=\Sigma^2
\delta(t-t')
\eeq
As to the nature of $\eta_t$ further complications are discussed in
\cite{ilinski:99a}, but they are not important for our analysis.
Basically, a stochastic component $x_t$ as been added to the constant
risk-free rate $r_0$. 
The question now is: How does the process for
the arbitrage return $x_t$ affect the price of a derivative ?

Substituting Eq.(\ref{arb.0}) 
for risk-free rate $r$ in the standard Black-Scholes PDE, Ilinski and
Stepanenko simply proceed and derive the following PDE:
\beq
\label{arb.1}
{\cal L}_{BS}V=x_t\left(V-S\pabl{V}{S}\right)
\eeq
where ${\cal L}_{BS}$ is the operator from the standard Black-Scholes
PDE, ${\cal L}_{BS}V=0$, for $r=r^0$.
We will clarify below that the replacement $r\rightarrow r^0+x_t$ in
fact is equivalent to introducing an interest rate process $r_t=r^0+x_t$
in a virtual world where tradable instruments dependant on this
interest rate exist.

The specific origin of intermediate arbitrage returns and market
incompleteness is assumed to be contained in the parameters $\lambda$
which sets the time scale for deviations from market equilibrium and
$\Sigma$ which gives a measure for the arbitrage returns
themselves. Specific values are discussed in section 6. 
Of
course, in reality transaction costs which are neglected here for simplicity might
effectively destroy small arbitrage returns.

In the next section, a different approach is presented in the
framework of standard option pricing theory which allows to study the
influence of intermediate deviations from financial equilibrium (as
defined by the no-arbitrage assumption) on derivative pricing.

\section{Derivatives in the presence of arbitrage opportunities: a mapping to 
option pricing theory with stochastic interest rates}
The way the arbitrage return $x_t$ has been introduced in the last section, in particular 
that the portfolio $\Pi_t$ grows at the rate $r^0+x_t$, allows for a mapping to option pricing theory with 
stochastic interest rates. We will call $r_t=r^0+x_t$ an interest rate in a virtual world, but do not insinuate 
that the arbitrage return is a part of the real interest rate. This virtual world will serve as a stage where 
known results can be used, but finally these results need to be projected to the real world. Let us 
for the moment assume, that this virtual world can be set up. The justification for its use will be delayed to 
section 5.

Let us first review the PDE approach to option pricing with stochastic interest rates.
The stochastic nature of a (short) interest rate $r_t$ is usually taken into
account by stating a stochastic differential equation (SDE) as follows:
\beq
\label{r}
dr_t=\rho(r_t,t) dt + \Sigma(r_t,t) dW^2_t
\eeq 
The parameters $\rho$ and $\Sigma$ specify drift and volatility
respectively, and may depend on $r_t$ and $t$.  The drift specifies
the deterministic (``trend'') component of the interest rate dynamics
whereas the volatility describes the stochastic fluctuations. The increment $dW^2_t$ is a
Wiener process.
The general PDE for a derivative $V=V(S,r,t)$ dependant on $S=S_t$ and $r=r_t$ can be found
in the literature \cite{hull:97}. Assuming for simplicity no
correlations between the Wiener processes $dW^1$ from Eq.(\ref{S}) and
$dW^2$ and suppressing the functional dependance of $\Sigma$ and $\rho$, the PDE is given by:
\beq
\pabl{V}{t}+\frac{\sigma^2 S^2}{2}\pabl{^2 V}{S^2}+S\pabl{V}{S}(\mu-\lambda_1\sigma)
-rV
+\frac{\Sigma}{2}\pabl{^2 V}{r^2}
+\pabl{V}{r}(\rho-\lambda_2 \Sigma)
=0
\eeq
The parameters $\lambda_i$, $i=1,2$ are known as the market prices of
risk for the security $S$ and the risk-free rate $r$. They can be
obtained by finding the change of measure which makes the respective
discounted price process a martingale \cite{baxter:96}.
For a
non-dividend paying security governed by Eq.(\ref{S}),
$\lambda_1=(\mu-r)/\sigma$. 
Incidentally, if $r$ is constant, one recovers the Black-Scholes PDE Eq.(\ref{BS}).
As $r$ is not a tradable security,
a tradable interest rate instrument is needed, e.g. a zero bond with maturity $T$ whose price at
time $t$ is $P(t,T)$ and which promises to pay one monetary unit at
time $T$. In fact, one is left with a residual freedom of
choosing $\lambda_2$ \cite{rebonato:96}. We will return to this issue
instantly.
Let us now restrict the drift $\rho_r$ to a mean-reverting form:
\beq
\label{vas}
\rho=\rho(r_t)=a-\lambda r_t
\eeq
Moreover, let us suppose that $\Sigma(r_t,t)=\Sigma=const$.
Let us further assume that 
\beq
\label{rplus}
r_t=r^0+x_t
\eeq
Then after transforming from $r$ to $x$ the PDE for the derivative price $V$ reads as:
\beq
\label{r.S}
\pabl{V}{t}+\frac{\sigma^2 S^2}{2}\pabl{^2 V}{S^2}+S\pabl{V}{S}
(r^0+x)-(r^0+x)V
+\frac{\Sigma^2}{2}\pabl{^2 V}{x^2}
+\pabl{V}{x}(a-\lambda r^0-\lambda_2 \Sigma-\lambda x)
=0
\eeq
As pointed out in \cite{rebonato:96}, it is not possible to separate
the market price of risk $\lambda_2$ from the difference $\tilde{a}=a-\lambda_2 \Sigma$.
Let us now return to the process for the arbitrage return $x_t$ assumed
in the last section, Eq.(\ref{arb}). Under the martingale measure for
the discounted zero bond price, the process for $x_t$ obtained from 
Eq.(\ref{r}) together with (\ref{vas}) and (\ref{rplus}) using standard
techniques \cite{baxter:96} is given by the SDE:
\beq
\label{x.Q}
dx_t=\left(\tilde{a}-\lambda r^0  -\lambda x_t\right)dt 
+ \Sigma d\tilde{W}^2_t
\eeq 
The change to a measure $\tilde{W}^2_t$ that makes the discounted zero bond price a martingale of 
course amounts to a choice as there is no such instrument in the real world. 
This means that there is no unique martingale measure in terms of real world instruments so we are necessarily forced 
to choose one.
One possibility is to require that the process
Eq.(\ref{x.Q}) under the martingale measure is mean-reverting to zero: deviations from economic equilibrium 
should disappear to zero. 
This means that
$\tilde{a}-\lambda r^0 = 0$, making Eq.(\ref{x.Q}) identical to 
the corresponding expression in
Eq.(\ref{arb}). 

Then
Eq.(\ref{r.S}) has a similar but not the same form as Eq.(3) in
\cite{ilinski:99b} (the last term should read $-\lambda x\pabl{V}{x}$
instead of $+\lambda \pabl{xV}{x}$),
\beq
\label{r.S.2}
\pabl{V}{t}+\frac{\sigma^2 S^2}{2}\pabl{^2 V}{S^2}+S\pabl{V}{S}
(r^0+x)-(r^0+x)V
+\frac{\Sigma^2}{2}\pabl{^2 V}{x^2}
-\lambda x\pabl{V}{x}
=0
\eeq
and the process for the arbitrage return becomes
\beq
\label{x.mart}
dx_t=-\lambda x_tdt 
+ \Sigma d\tilde{W}^2_t
\eeq
yielding the consistency of the no-arbitrage approach in the virtual
world with the arbitrage dynamics proposed in Eq.(\ref{arb}).
The security price dynamics becomes
\beq
dS_t=(r^0+x_t) S_tdt + \sigma S_t d\tilde{W}^1_t
\eeq
with respect to the martingale measure.This equation basically couples arbitrage returns to 
security price dynamics under the chosen martingale measure. Thus incompleteness is introduced here in 
terms of stochastic drift as mentioned above.

Of course, the dynamics of real interest rates are not the same as the
dynamics proposed here for a virtual world. The relaxational time scale
$1/\lambda$ originating from the disappearance of virtual arbitrage
returns is much shorter than a time scale of mean reversion for real
interest rates. As will become clear in the next section, our zero
bond price in the virtual world will approach a real (constant interest rate) bond price in
the limit of infinitely fast relaxation dynamics for the arbitrage
return $x_t$. One might object at this point, that we have assumed a
hedging strategy in the virtual world which does not exist in the real
world ($x_t$ cannot 
be hedged). Indeed, the hedging strategy in the virtual world
expressed in terms of the security $S_t$ and a real world cash bond
$B^0_t=\exp(-r^0t)$ leaves us with an extra amount arising from the
dynamics of $x_t$, and is therefore not {\it self-financing} in terms
of real world instruments.  We will address this issue in more detail in section 5.

In order to complete the pricing problem in the virtual world,
Eq.(\ref{r.S.2}) requires the boundary condition, e.g. for a European claim
(which is exercised or not exactly at time $T$). It must be chosen as \cite{ilinski:99b}
\beq
\label{payoff}
V(t,S,r)|_{t=T}=X\delta(x)
\eeq
where $X$ is the final payoff (in the real world) depending on
$S(T)$ and $r=r^0+x$. 
Then the option price in the real world may be calculated as an average over the
initial arbitrage return as follows:
\beq
\label{V.av.0}
\bar{V}(t,S,r^0)=\int_{-\infty}^\infty dx V(t,S,r) \tilde{p}(x)
\eeq
where $\tilde{p}(x)$ is a probability density function chosen
according to the dynamics of $x_t$ to be discussed below.

However, we will not proceed to solve the PDE, but remember that
according to the Feynman-Kac lemma \cite{feynman:48,kac:49}
\beq
V(t,S,r)=E_{\rm Q}\left[e^{-\int_t^T ds r_s}X\delta(x_T)|x_t=x,r^0, S_t\right]
\eeq
Now it is easy to show (e.g. by using the path integral approach \cite{otto:98}) that
\beqa
E_{\rm Q}\left[e^{-\int_t^T ds r_s}X\delta(x_T)|x_t=x,r^0, S_t\right]
&=&E_{\rm Q}\left[e^{-\int_t^T ds r_s}X|x_t=x,r^0, S_t;x_T=0\right]
\nonumber\\
&\times&p(x_T=0|x_t=x,r^0, S_t)
\eeqa
where $p(x_T=0|x_t=x,r^0, S_t)$ is the conditional probability density
function for $x_T=0$ given $x_t=x$,$r^0$,$S_t$.
The last equation allows us to utilize results from the literature on option pricing
theory with stochastic interest rates. 
In fact, we will solve the pricing problem for the actual payoff
function $X$ in an interest rate environment where the short rate
process in the virtual world $r_t$ starts from $r^0+x$ at time $t$ and comes
back to $r^0$ at the time of maturity $T$, or put otherwise where
$x_t=x$ and $x_T=0$.

The average as given in Eq.(\ref{V.av.0}) will then be performed in a
different way.
From the constraint $\bar{V}(t,S,r_0)|_{t=T}=X$, it clear that 
\beq
\tilde{p}(x)=\frac{p(x)}{p(0)}
\eeq
where $p(x)$ is the probability density function for the initial value
$x$ of the arbitrage return.
Using the fact that in our case $p(x_T=0|x_t=x,r_0,
S_t)=p(x_T=0|x_t=x)$ and
\beq
p(x_t=x|x_T=0)=p(x_T=0|x_t=x)\frac{p(x)}{p(0)}
\eeq
one may rewrite the average in Eq.(\ref{V.av.0}) as follows:
\beq
\label{V.av}
\bar{V}(t,S,r^0)=\int_{-\infty}^\infty dx V(t,S,r;x_T=0) p(x_t=x|x_T=0)
\eeq
where 
\beq
V(t,S,r;x_T=0)=E_{\rm Q}\left[e^{-\int_t^T ds r_s}X|x_t=x,r^0, S_t;x_T=0\right]
\eeq
and
where $p(x_t=x|x_T=0)$ is the conditional probability density for
the arbitrage return at time $t$ to be equal to $x$ given that its
value at $T>t$, the time of expiry, is zero. Its explicit form will be
discussed in the next section. In fact, as $t=T$, one obtains
$p(x_t=x|x_T=0)=\delta(x)$ as required. 

\section{Valuation of European call and put options in the presence of virtual arbitrage opportunities}
Considering the arbitrage return as a part of the stochastic interest
rate $r_t$ in our virtual world, we can now draw upon a classical result
of Merton \cite{merton:73} in order to derive formulas for European
call and put options. In fact, instead of solving
Eq.(\ref{r.S.2}) together with Eq.(\ref{payoff}) for $X=\max(S-K;0)$ (or
$X=\max(K-S;0)$) for call or put options respectively,  we consider
the security price dynamics Eq.(\ref{S}) together with the following SDE
for 
the price of a zero bond:
\beq
\label{bond}
d_t P(t,T)=P(t,T)
\left(
\mu_P(t,T) dt +\sigma_P(t,T)dW^2_t
\right)
\eeq
Now assuming that $\sigma_P(t,T)$ is a known function of $t$ and $T$, the price of a
European call (and put) option $c$ (and $p$) at time $t$ which
expires at time $T$ with strike price $K$ is given by \cite{merton:73}:
\beqa
\label{merton}
c&=&SN(d_1)-P(t,T)KN(d_2)\\
p&=&P(t,T)KN(-d_2)-SN(-d_1)
\eeqa
where $N(x)$ is the cumulative normal distribution and 
\beqa
\label{param}
d_1&=&\frac{\ln(S/K)-\ln(P(t,T))+\hat{\sigma}^2(T-t)/2}{\hat{\sigma}\sqrt{T-t}}
\nonumber\\
d_2&=&d_1-\hat{\sigma}\sqrt{T-t}\nonumber\\
\hat{\sigma}^2(T-t)&=&\int_t^T ds
\left(
\sigma^2+\sigma_P^2(s,T)-2\rho_{SP} \sigma\sigma_P(s,T)
\right)
\eeqa
The parameter $\sigma$ is the volatility of the security, and $\rho_{SP}$ is
the instantaneous correlation between the stock and zero bond prices
which for simplicity we assume to be zero as above. Let us now connect to
the stochastic interest rate dynamics $r_t$. 
If the process for $P(t,T)$ is derived from the process for $r_t$
using Ito's lemma, one obtains the following dependance of the bond
volatility on the parameters of the process for $r_t$:
\beq
\label{bondvola}
\sigma_P(t,T)=\Sigma\frac{1}{P(t,T)}\pabl{P}{r}
\eeq
where $\Sigma$ is the short rate volatility. For the specific short
rate dynamics chosen in Eq.(\ref{r}) and (\ref{vas}), the zero bond
price can be calculated explicitly as a function of $t$, $T$, the
current short rate level $r$ and the model parameters. Concerning the latter
ones, the drift of the short rate process has to be risk-adjusted by
the market price of risk giving $\tilde{a}$ as mentioned above.
Let us now calculate the bond price $P(t,T)$. Using the fact that
\beqa
P(t,T)&=&E_{\rm Q}\left[e^{-\int_t^Tds r_s}|r_0\right]\nonumber\\
&=&e^{-r^0(T-t)}E_{\rm Q}\left[e^{-\int_t^Tds x_s}|x_t=x\right]
\eeqa
one obtains from the dynamics of $x_t$ Eq.(\ref{x.mart}) (after a tedious calculation given in the appendix using the
path integral approach \cite{otto:98}) the following result:
\beq
\label{P.res}
P(t,T)=\exp\left(
-\left(r^0-\frac{\Sigma^2}{2\lambda^2}\right)(T-t)-
\frac{1}{\lambda}\tanh\left(\frac{\lambda (T-t)}{2}\right)\left(x+\frac{\Sigma^2}{\lambda^2}\right)
\right)
\eeq
Performing the differentiation in Eq.(\ref{bondvola}), one obtains:
\beq
\label{bondvola2}
\sigma_P(t,T)=-\frac{\Sigma}{\lambda}\tanh\left(\frac{\lambda (T-t)}{2}\right)
\eeq
Now, the restriction $\tilde{a}=\lambda r^0$ which makes the drift of the process
for the arbitrage return $x_t$ be equal to $-\lambda x_t$ (under the
martingale measure), gives the desired asymptotics of the zero bond price.
In fact, 
as $\lambda\rightarrow \infty$, which can be interpreted as an infinitely
fast disappearance of virtual arbitrage returns, it reads as
\beq
\label{asymp1}
\lim_{\lambda\rightarrow \infty}P(t,T)=e^{-r^0(T-t)},
\eeq
the zero bond price for a constant risk-free rate $r^0$.
Therefore the restriction on $\tilde{a}$ mentioned above, and thus our choice of the martingale measure 
 is reasonable also from the viewpoint of correct zero bond price asymptotics.
Let us know turn to the evaluation of the modified security price
volatility $\hat{\sigma}$. Evaluating the integral in Eq.(\ref{param}), one obtains:
\beq
\hat{\sigma}^2=\sigma^2+\frac{\Sigma^2}{\lambda^2}
\left(
1-\frac{2}{\lambda(T-t)}\tanh\left(\frac{\lambda (T-t)}{2}\right)
\right)
\eeq
Likewise, in the limit $\lambda\rightarrow \infty$, the contribution
to virtual arbitrage returns disappears and one recovers the ``bare''
security price volatility:
\beq
\label{asymp2}
\lim_{\lambda\rightarrow \infty}\hat{\sigma}=\sigma
\eeq
The asymptotic equations Eq.(\ref{asymp1}) and (\ref{asymp2}) assure
that in the case of infinitely fast vanishing arbitrage returns the
Black-Scholes formulas (for a constant risk-free rate $r^0$) are
recovered from Eq.(\ref{merton}).
When the option approaches maturity, there is the following expansion of
$\hat{\sigma}^2$:
\beq
\hat{\sigma}^2=\sigma^2+\frac{\Sigma^2}{12}
(T-t)^2+{\cal O}\left((T-t)^3\right)
\eeq
Now as the option price in our virtual world is fixed in terms of the
parameters of the arbitrage return process, we need to turn to the
explicit evaluation of the average carried out in
Eq.(\ref{V.av}). For an Ornstein-Uhlenbeck process  $x_t$  given by Eq.(\ref{x.mart}) it is well
known (e.g.\cite{feller:68}), that the transition probability to go from $x'$ at time $0$ to
$x$ at time $t$ is given by:
\beq
\label{o-u}
p(x_t=x|x_0=x')=\sqrt{\frac{\lambda}{\pi \Sigma^2}}
\left(1-e^{-2\lambda t}\right)^{-1/2}
\exp\left(
-\frac{\lambda}{\Sigma^2}
\frac{\left(x-x'e^{-\lambda t}\right)^2}{\left(1-e^{-2\lambda t}\right)}
\right)
\eeq
What is needed however in our case, is $p(x_t=x|x_T=0)$ for $T\geq
t$ 
which is
obtained from Eq.(\ref{o-u}) as follows:
\beq
p(x_t=x|x_T=0)=p(x_T=0|x_t=x)\frac{p(x)}{p(0)}
\eeq
where $p(x)$ is the probability density for $x$ which is obtained as a
limit probability density form Eq.(\ref{o-u}) as $t\rightarrow
\infty$:
\beq
p(x)=\sqrt{\frac{\lambda}{\pi \Sigma^2}}\exp\left(
-\frac{\lambda}{\Sigma^2}
x^2
\right)
\eeq
The final expression for the transition probability thus reads:
\beq
\label{o-u2}
p(x_t=x|x_T=0)=\sqrt{\frac{\lambda}{\pi \Sigma^2}}
\left(1-e^{-2\lambda (T-t)}\right)^{-1/2}
\exp\left(
-\frac{\lambda}{\Sigma^2}x^2
\frac{1}{\left(1-e^{-2\lambda (T-t)}\right)}
\right)
\eeq
It has all the desired features needed. Using the following
representation of Dirac's delta function:
\beq
\lim_{n\rightarrow\infty}ne^{-\pi n^2 x^2}=\delta(x)
\eeq
one obtains both for the limit of infinitely rapid disappearance of arbitrage
returns
\beq
\lim_{\lambda/\Sigma^2\rightarrow \infty}p(x_t=x|x_T=0)=\delta(x)
\eeq
and for the limit $t\rightarrow T$ of approaching the option's time of maturity 
\beq
\lim_{t\rightarrow T}p(x_t=x|x_T=0)=\delta(x)
\eeq
In both cases, one expects arbitrage returns to disappear.
Next, the average over virtual arbitrage returns in Eq.(\ref{V.av}) is
carried out explicitly for a European call option (for $\bar{V}=\bar{c}$) as
\beqa
\label{c.warb}
\bar{c}(t,S,r_0)&=&S\int_{-\infty}^\infty dx 
N(d_1)
p(x(t)=x|x(T)=0)\nonumber\\
&-&K\int_{-\infty}^\infty dx P(t,T)N(d_2)
p(x(t)=x|x(T)=0)
\eeqa
and a European put option (for $\bar{V}=\bar{p}$) as
\beqa
\label{p.warb}
\bar{p}(t,S,r_0)&=&K\int_{-\infty}^\infty dx P(t,T)
N(-d_2)
p(x(t)=x|x(T)=0)\nonumber\\
&-&S\int_{-\infty}^\infty dx N(-d_1)
p(x(t)=x|x(T)=0)
\eeqa
where $P(t,T)$ is given in Eq.(\ref{P.res}).
The integrations with respect to $x$ cannot be performed
analytically. 
However, the integrands decrease sufficiently fast to zero as
$x\rightarrow\pm\infty$, so that a numerical integration can be
 be easily performed. 

It is obvious from intuition that the pricing formulas Eq.s
(\ref{c.warb}) and (\ref{p.warb}) contain the fundamental time scale
$\tau_{\rm arbitrage}=1/\lambda$. In fact, one can introduce the
following scaled variables:
\beqa
\label{scaled}
u&=&\lambda(T-t)=\frac{(T-t)}{\tau_{\rm arbitrage}}\nonumber\\
r_\lambda&=&\frac{r}{\lambda}\nonumber\\
x_\lambda&=&\frac{x}{\lambda}\nonumber\\
\hat{\sigma}_\lambda&=&\frac{\hat{\sigma}}{\sqrt{\lambda}}\nonumber\\
\sigma_\lambda&=&\frac{\sigma}{\sqrt{\lambda}}\nonumber\\
\Sigma_\lambda&=&\frac{\Sigma}{\lambda^{3/2}}
\eeqa
Then $\lambda$ can be eliminated from the pricing formulas. 
The parameters in Eq.(\ref{param}) can expressed in terms of the
scaled variables of Eq.(\ref{scaled}):
\beqa
\label{scaled.p}
d_1&=&\frac{\ln(S/K)-\ln(P(u))+\hat{\sigma}_\lambda^2u/2}{\hat{\sigma}_\lambda\sqrt{u}}
\nonumber\\
d_2&=&d_1-\hat{\sigma}_\lambda\sqrt{u}\nonumber\\
\hat{\sigma}_\lambda^2&=&\sigma_\lambda^2+\Sigma_\lambda^2
\left(
1-\frac{2}{u}\tanh\left(\frac{u}{2}\right)
\right)
\eeqa
where $P(u)$ is given by:
\beq
P(u)=P(t,T)=\exp\left(
-\left(r^0_\lambda-\frac{1}{2}\Sigma_\lambda^2\right)u-
\tanh\left(\frac{u}{2}\right)\left(x_\lambda+\Sigma_\lambda^2\right)
\right)
\eeq

\section{Replicating hedging strategies}
The issue of hedging strategies in the virtual and
the real world
mentioned above
will now be addressed. 
The fact that there is no instrument in the real world to hedge
intermediate arbitrage returns leads us to conjecture that a hedging
strategy might not be self-financing.

To be specific, let us denote a
cash bond in our virtual world as follows:
\beq
\label{cash.v}
B_t=\exp\left(\int_0^t ds r_s\right)
\eeq
It monitors the temporal evolution of the value of an initial cash deposit $B_0=1$
which earns the instantaneous interest rate $r_s$. Let us further
introduce the cash bond in the real world
\beq
\label{cash.r}
B^0_t=\exp(r^0t)
\eeq
and as a further abbreviation ( which may be termed the ``arbitrage bond'') 
\beq
\label{cash.a}
B^x_t=\exp\left(\int_0^t ds x_s\right)
\eeq
Evidently, one obtains:
\beq
\label{v2r}
B_t=B^0_t B^x_t
\eeq
Taking $B_t$ for the moment as a real cash bond, a {\it self-financing}
strategy $V_t$ consists of holding $\varphi_t$ in the security $S_t$ and
$\psi_t$ in the cash bond $B_t$ such that
\beq
\label{sf}
V_t=\varphi_t S_t+\psi_t B_t \Rightarrow dV_t=\varphi_t dS_t+\psi_t dB_t
\eeq
i.e. the value change $dV_t$ is only due to price changes $dS_t$ and
$dB_t$.
For our security price model with stochastic interest rates in the
virtual world, one can show that Eq.(\ref{sf}) holds \cite{baxter:96}. Moreover
$V_T=X$, i.e. the value of portfolio $V$ equals the final payoff, i.e. it is {\it replicating}. So in terms of
our fictitious cash bond $B_t$ there is a self-financing, replicating
strategy. In the real world, our strategy will remain replicating by
construction (see Eq.s(\ref{V.av.0}) to (\ref{V.av})). However, it
will not be self-financing in terms of the real cash bond $B^0_t$ and
the security price $S_t$, as can be seen by substituting for $B_t$ in
Eq.(\ref{sf}):
\beqa
\label{cost}
dV_t&=&\varphi_t dS_t+\psi_t d(B^0_tB^x_t)\nonumber\\
&=&\varphi_t  dS_t+\psi_t B^x_t dB^0_t +\psi_t B^0_t dB^x_t\nonumber\\
&=&\varphi_t  dS_t+\psi_t B^x_t dB^0_t +\psi_t B^0_t B^x_t x_t
dt\nonumber\\
&=&\varphi_t  dS_t+\psi_t B^x_t dB^0_t+(V_t-\varphi_tS_t)x_t dt
\eeqa
The last step was to replace $\psi_t B^0_t B^x_t=\psi_t B_t$ by
$V_t-\varphi_tS_t$ using Eq.(\ref{sf}).
The third term on the r.h.s of the last line 
accounts for extra
costs or gains due to arbitrage opportunities.
It is exactly equal to the instantaneous (positive or negative) arbitrage
return earned on the delta hedge $V_t-\varphi_tS_t$. In fact, one has
$\Delta=\varphi_t$, and therefore
\beq
\Pi_t=V_t-\varphi_tS_t
\eeq
where $\Pi_t$ is the delta hedge portfolio discussed in section 2. The
replacement $r\rightarrow r^0+x_t$ introduced by Ilinski
\cite{ilinski:99a} gives rise to the same additional term in the hedging
strategy $V_t$, if one considers the change $d\Pi_t$ as follows:
\beq
d\Pi_t=(r^0+x_t)\Pi_tdt=r^0\Pi_t dt+x_t\Pi_t dt
\eeq
The second term on the r.h.s of this equation is the source of
additional intermediate profit and loss (p\&l) during the hedging process. Therefore, we conclude that the
replacement of Ilinski is completely equivalent to the introduction of
a fictitious cash bond $B_t$ or likewise an interest rate $r_t$ as defined above, which ensures a
self-financing hedging strategy in the virtual world.

The additional hedging costs or gains which arise in the real world are covered
by an additional premium contained in the option price as obtained in
Eq.(\ref{V.av}) (with respect
to the Black-Scholes price). This premium
is positive in most cases as will be clarified below when numerical
examples are discussed.

Finally, let us further back up the interpretation of
$(V_t-\varphi_tS_t)x_tdt$ as representing the differential p\&l  on
the hedging strategy within the time interval $dt$, using the following
argument (whose formulation is borrowed from
\cite{bouchaud:97}). 
At time $t$, an option is sold at $O_t$ in the real world, and using
the premium the following portfolio $\langle V_t\rangle$ is bought:
\beq
\langle V_t\rangle=\langle\varphi_t\rangle
S_t+\langle\psi_tB^x_t\rangle B^0_t=O_t
\eeq
where $\langle\dots\rangle$ corresponds to an average over all paths
$\{x_s\}_{s\in[t,T]}$.
Furthermore, the change in wealth of the option seller within the time
interval $[t,T]$ in the real world is given by:
\beq
\Delta W=O_t+\int_t^T\langle\varphi_s\rangle dS_s+
\int_t^T \langle V_s-\varphi_sS_s\rangle r^0 ds
+\int_t^T\langle V_s-\varphi_sS_s\rangle x_s ds-X
\eeq
The first term is the option premium earned, the second term gives
the cumulative gain by the trading the asset, the third one corresponds to the
cost/gain of the cash bond position (used to finance the position in
the asset or set side as excess cash respectively) which is
proportional to the riskless rate $r^0$, and the fourth
term is supposed to take into account the p\&l due to virtual
arbitrage. In fact, the fourth term can be added to third
term giving an effective cost/gain of the cash bond position due to
the effective rate $r^0+x_t$.
Finally the last term is the potential cash outflow due to the
option's payoff.
Now using $d\langle V_t\rangle=\langle\varphi_t\rangle
dS_t+\langle(V_t-\varphi_tS_t)(r^0+x_t)\rangle dt
$ one shows that:
\beq
\Delta W=O_t+\int_t^Td\langle V_s\rangle-X=O_t-\langle
V_T\rangle-\langle V_t\rangle-X=0
\eeq
as $V_T=X$ by construction. As $\Delta W$ vanishes identically,
$\overline{\Delta W^2}(t,S,r^0)$ (where the average is taken as in Eq.(\ref{V.av}))) vanishes as well which implies that
no intrinsic risk remains over the remaining time to maturity of the option, and therefore no risk-minimization is necessary. The influence of virtual arbitrage is
completely taken care of by the option premium. We see also that our
hedging strategy in the real world is not self-financing at every time
step but is self-financing when the time integral over remaining life
time of the option is taken. 

As discussed e.g. in \cite{foelmer:91}, incomplete markets imply that
there is no unique equivalent martingale measure any more. However,
martingale theory may still be used if a supplementary constraint is
added (see the discussion presented in the introduction)
 which then selects a particular martingale measure. In our case this choice has been implicitly made when the
arbitrage return becomes part of a fictitious interest rate in the
virtual world. In fact, both local risk (expected conditional squared cost) and
replication risk (expected squared deviation of the terminal hedging portfolio
to payoff) \cite{heath:98} are trivially minimized, i.e. zero. A
detailed comparison of our approach to incomplete markets to the
F\"ollmer-Schweizer approach \cite{foelmer:91} certainly deserves further study.

\section{Some numerical results}
In the following, some results are presented for two market
situations, a rather incomplete market (FIG.(\ref{call1}) and
(\ref{put1})) and a fairly complete market (FIG.(\ref{dc1}) and
(\ref{dp1})). In the first case, the averaged prices $\bar{c}$ (and $\bar{p}$) , the
Black-Scholes prices and the payoff functions at maturity are given, for
parameters 
$\lambda=10$,
$T-t=0.8$, $\Sigma=2$, $\sigma=0.2$, $K=100$,
$r^0=0.08$. The unit of time is 1 year, so $\lambda=10$ corresponds to
the rather long relaxation time
$\tau_{\rm arbitrage}$  of about $25$ trading days, supposing a year of
$250$ trading days. $\Sigma=2$ is inferred from a daily maximum
variation of $x_t$ of about $20\%$ in absolute value (at $95\%$ confidence level)
according to the discretized Langevin equation:
\beq
\Delta x=x_{t+1}-x_t=-\lambda x_t \Delta t+X\Sigma\sqrt{\Delta t}
\eeq
The random variable $X$ is standard normally distributed. Taking
$\Delta t=1/250$, $X=1.65$ representing the two-sided $95\%$ 
confidence interval, $x_t=0$ (as an initial value), one concludes
\beq
\Sigma=9.58\Delta x
\eeq
For various times to maturity $T-t$,
FIG.(\ref{dc1}) and
(\ref{dp1}) presents differences of
$\bar{c}$ (and $\bar{p}$) and the Black-Scholes prices for the choice
of parameters $\lambda=100$,
$\Sigma=0.4$, $\sigma=0.2$, $K=100$,
$r^0=0.08$. $\lambda=100$ corresponds to a relaxation time of $2$ to
$3$ trading days, whereas $\Sigma=0.4$ is inferred from a daily
variation of $x_t$ of about $4\%$ in absolute value. 

Let us now comment on the results. Focusing first on the qualitative
behavior, 
over a reasonable range of the moneyness parameter $m=S/K$, the price
of a European call or put option ($\bar{c}$ or $\bar{p}$ respectively) under the influence of virtual
arbitrage is higher than the Black-Scholes value (see
FIG.(\ref{call1}) and (\ref{put1})). The
difference is more pronounced at the point of maximum curvature which
is around $m\simeq 1$ or below , whereas it decreases
whenever $m<1$ or $m>1$ (see FIG.(\ref{dc1}) and (\ref{dp1})). For $m\gg 1$, the call option price is less than
the Black-Scholes value. As the time to expiry increases the positive
difference (except for $m\gg 1$) increases, and the maximum difference
is shifted to lower values of $m$. 

Leaving aside for the moment the negative difference appearing for
call option at $m\gg 1$, it appears reasonable that the existence of virtual arbitrage returns
causes the option price to be above the Black-Scholes value, as deviations
from equilibrium in general lead to an increase in hedging costs, i.e. the costs for
readjusting a replication portfolio which is supposed to provide for
the final payoff of the option. This effect needs to be accounted for
in the option premium. 
The fact that the absolute difference to the Black-Scholes result is
the largest at the point of maximum curvature of the pricing function
is understandable from the $\Gamma$ (``gamma'') risk point of
view. $\Gamma$ denotes the second derivative of the option price with
respect the asset price $S$ and gives a measure for the non-linear
dependance of the option on the underlying asset. This non-linear risk
inherent to options can be only be hedged by buying or selling other
options. Any deviations from financial equilibrium due to arbitrage
opportunities will affect both the option at hand and the options
chosen for hedging. Moreover, the additional term arising in
Eq.(\ref{cost}) leading to intermediate P\&L during the hedging
process is proportional to the delta hedge $V_t-\varphi_tS_t$ which is
most relevant at the point of maximum $\Gamma$ where the delta hedge
is insufficient. Therefore, these numerical results are completely
consistent with our mathematical discussion of the hedging strategy.

The influence of intermediate arbitrage returns grows as the time to
expiry of the option increases on the scale of $\tau_{\rm
  arbitrage}$ (see FIG.(\ref{dc1}) and (\ref{dp1})) (all other parameters being constant).
 Several deviations from equilibrium during the life
time of an option seem to accumulate leading to a higher
additional risk premium on the
option price. 

Returning to the issue of the negative difference for call options
that are far in the money $m\gg 1$ in FIG.(\ref{call1}) and (\ref{dc1}), a possible explanation is an
``overheated'' market, where deviations from equilibrium tend to relax
from the current asset price to a lower equilibrium
price. This information is accounted for by pricing the option at a
discount with respect to the Black-Scholes value at the current asset
price: the market is expected to decrease to a lower price level.

Considering the quantitative differences between $\bar{c}$ (and
$\bar{p}$) and the Black-Scholes prices for calls and puts, they are
obviously more pronounced in an imcomplete market (FIG.(\ref{call1})
and (\ref{put1})), than in a rather complete market where arbitrage
returns are small and relax fast ((\ref{dc1}) and (\ref{dp1})). The
numerical analysis given here may be refined in various ways (according to the parameter
dependances of the options prices) which is the subject of future work.

As opposed to our results, e.g. the first order correction to
the Black-Scholes prices for calls given in
\cite{ilinski:99a} increases monotonously with moneyness
$m$. Obviously, as our result is based on the same model as in
\cite{ilinski:99a} (see section 5) and is exact (apart from the
remaining integration over the initial arbitrage), some error is made
in the perturbative treatment. Let us point out here again that the
Ornstein-Uhlenbeck dynamics of arbitrage returns used here and the fact
that the $x_t$ gives the extra return on the delta hedge
$V_t-\varphi_tS_t$, it is quite reasonable that the difference to the
Black-Scholes price should show a maximum at the price level where the
delta hedge fails.

\section{Conclusion}
Using the (Ornstein-Uhlenbeck type) relaxational dynamics for ``virtual'' arbitrage returns
introduced in \cite{ilinski:99a}, we have derived closed formulas for
simple (``plain vanilla'') European calls and puts in the presence of
arbitrage opportunities appearing and disappearing on an intermediate
time scale $\tau_{\rm arbitrage}=1/\lambda$. This result which has not
been derived previously is obtained using martingale
option pricing theory for incomplete markets (in the sense of \cite{foelmer:91}), by making the arbitrage return process part of an 
interest rate process in a virtual world. The influence on option prices
in the real world (in the presence of rapidly appearing and
disappearing arbitrage opportunities) is taken into account by summing
over the initial arbitrage return, and imposing the constraint that
arbitrage is absent at the time of maturity of the option.

Comparing our work to \cite{ilinski:99a,ilinski:99b}, first, we consider the
analysis given above as conceptually more clear as to where
arbitrage-free pricing fails and where it does not.
Therefore, in the present work a different route has been
proposed by
 introducing a second source of
randomness in the derivative pricing problem (apart from the security
$S$) right from the beginning. As a consequence, a two variable
version of Ito's lemma must be used, giving a PDE equation for 
the derivative price in a virtual world which is finally summed over $x_t$
to yield the real world price.
Second, instead of making the constraint that arbitrage return should
vanish at maturity a part of the payoff function in the virtual world
as in \cite{ilinski:99a},
we enforce it when the average over virtual arbitrage return is
taken. This procedure allows us to profit from Merton's classical result on
option pricing in a stochastic interest rate environment
\cite{merton:73} and to arrive at closed-form (up to a numerical
integration over the initial arbitrage return which is easy to perform) pricing formulas for simple
European call and put options.

Furthermore it has been shown that any hedging strategy will not be
self-financing in the real world where the arbitrage return is not
directly observable. However, on the average any intermediate costs
arising during the hedging process are covered by an additional
premium contained in the option price. In this sense, a hedging
strategy can be found that is self-financing in a time average sense,
i.e. when summed over the remaing life-time of the option.
The derivation of pricing formulas rests crucially on the selection of
a specific measure from a set of equivalent martingale measures that
contains more than one element, due to intermediate market
incompleteness which arises because of virtual arbitrage opportunities.

The present work may be extended in various directions. The
relaxational dynamics of the arbitrage return may be considered to be
more complicated as proposed here where it follows a simple
Ornstein-Uhlenbeck process. However, additional model parameters
introduce more sources of model error from the practitioner`s point of view
as each parameter has to calibrated to the market.
Furthermore, the constraint $x_T=0$, i.e. that arbitage returns should disappear at the time of maturity 
of the option, may be relaxed to allow for a hedging mismatch at maturity. This amounts to give up 
the constraint that the hedging strategy is replicating. The extension
of this work to the case of correlations between the asset price $S_t$
and the arbitrage return $x_t$ is under way. Certainly, the comparison of the present model to stochastic volatility
models deserves further study.
\\
\\
{\bf Acknowledgements}: 
Discussions with M. Wilkens and participants of the "Internes Seminar Wertpapiermanagement" at 
the Institut f\"ur Betriebliche Geldwirt-\\schaft (IfBG), Universit\"at G\"ottingen 
are gratefully acknowledged.
The author thanks A.K. Hartmann for a critical
reading of the manuscript, and acknowledges financial support by the DFG under
grant Zi209/6-1.

\section*{Appendix}

Following \cite{otto:98} we propose to evaluate the expectation value
\beq
I=E_{\rm Q}\left[e^{-\int_t^Tds x_s}|x_t=x\right]
\eeq
The expectation value can be stated in terms of a quotient of path
integrals as follows:
\begin{equation}
\label{P1}
I=\frac{\int_{x(t)=x}^{x(T)=0}{\cal D}x(s)
\exp\left(-\frac{1}{2\Sigma^2}\int_t^T ds 
\left(\dabl{x(s)}{s}+\lambda x(s)\right)^2
-\int_t^T ds x(s)
\right)}{\int_{x(t)=x}^{x(T)=0}{\cal D}x(s)
\exp\left(-\frac{1}{2\Sigma^2}\int_t^T ds 
\left(\dabl{x(s)}{s}+\lambda x(s)\right)^2
\right)} =\frac{X}{Y}
\end{equation}
Now the numerator and the denominator can be mapped to the propagator
of the harmonic oscillator in the presence of an external field, and
can thus be evaluated \cite{feynman:72}. The expression for the numerator reads
as
\begin{eqnarray}
\label{num}
X&=&\sqrt{\frac{\lambda}{2\pi \Sigma^2 \sinh(a(T-t)}}
\exp\left(
\frac{\Sigma^2}{2\lambda^3}\left(
e^{-\lambda(T-t)}-1+\lambda(T-t)
\right)
\right.\nonumber\\
&-&\left.\frac{\lambda}{2\Sigma^2 \sinh(\lambda(T-t))}
\left(
x^2\cosh(\lambda(T-t))
\right.\right.\nonumber\\
&+&\left.\left.
2\left(e^{\lambda(T-t)}-1\right)(Cx+C^2)
\right)
\right.\nonumber\\
&+&\left.\frac{\lambda}{2\Sigma^2}
x^2
\right)
\end{eqnarray}
where 
\beq
C=\frac{\Sigma^2}{2\lambda^2}
\left(e^{-\lambda(T-t)}-1\right)
\eeq
Likewise one obtains an expression for the denominator:
\begin{eqnarray}
\label{denom}
Y&=&\sqrt{\frac{\lambda}{2\pi \Sigma^2 \sinh(\lambda(T-t)}}
\exp\left(
\frac{\lambda}{2\Sigma^2}
x^2
-\frac{\lambda}{2\Sigma^2 \sinh(\lambda(T-t))}
x^2\cosh(\lambda(T-t))
\right)\nonumber\\
\end{eqnarray}
Calculating $X/Y$ gives the result
\beq
I=\exp\left(
\frac{\Sigma^2}{2\lambda^2}(T-t)-
\frac{1}{\lambda}\tanh\left(\frac{\lambda (T-t)}{2}\right)\left(x+\frac{\Sigma^2}{\lambda^2}\right)
\right)
\eeq
which leads to Eq.(\ref{P.res}).

\begin{figure}
  \begin{center}
  \epsfig{file=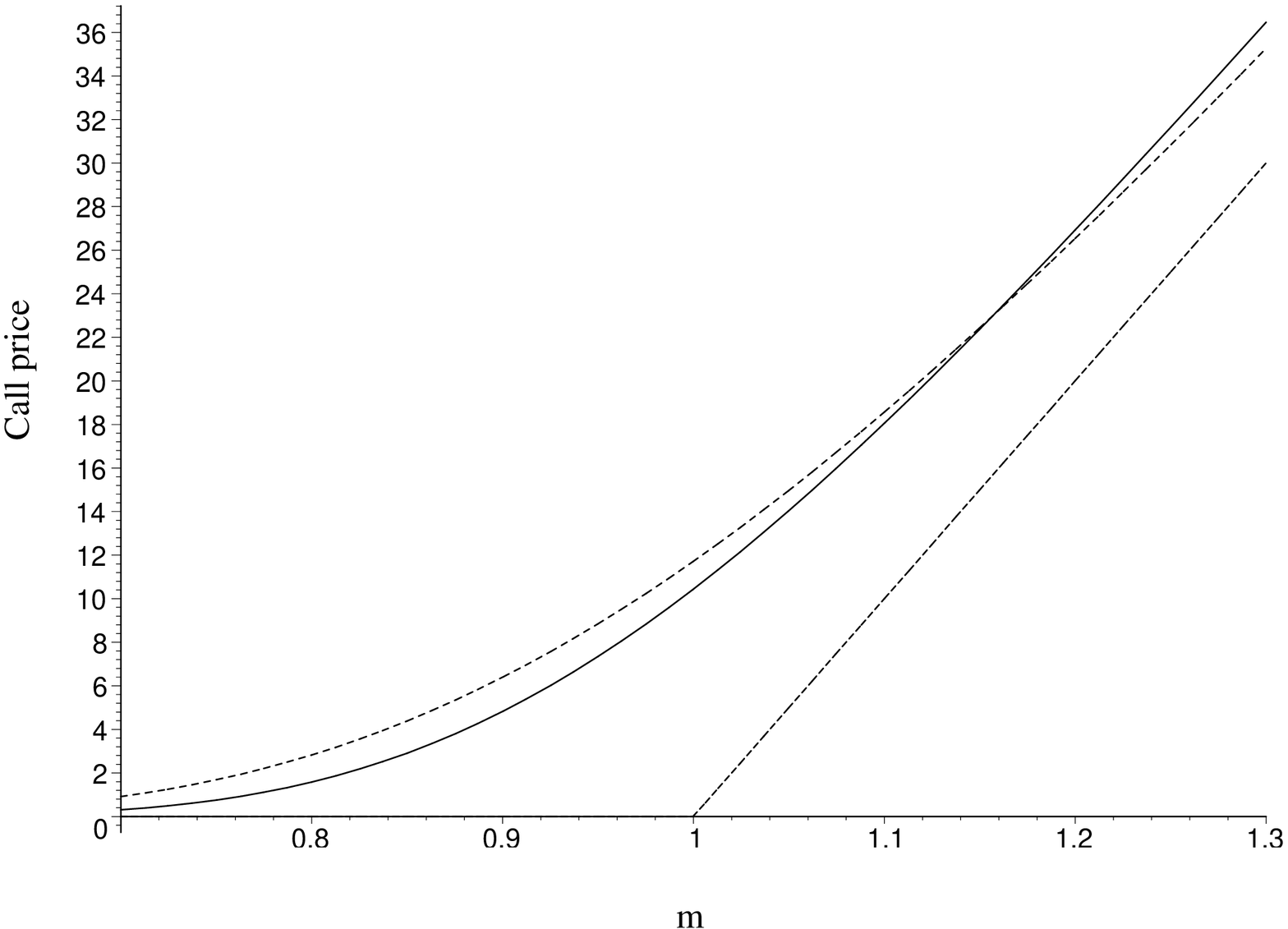,angle=90}
  \end{center}
  \caption{}
  \label{call1}
\end{figure}

\begin{figure}
  \begin{center}
  \epsfig{file=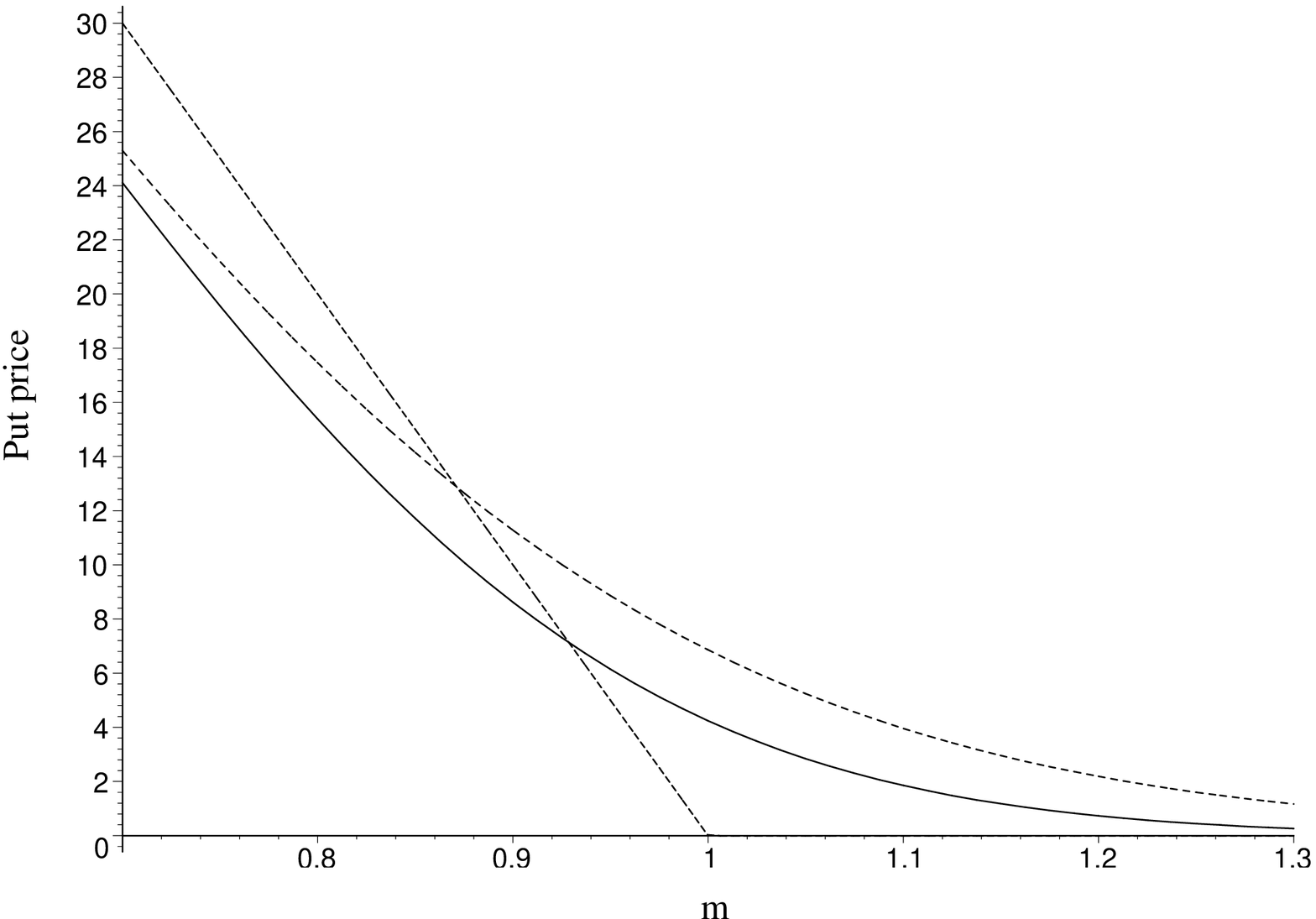,angle=90}
  \end{center}
  \caption{}
  \label{put1}
\end{figure}

\begin{figure}  
  \begin{center}   
  \epsfig{file=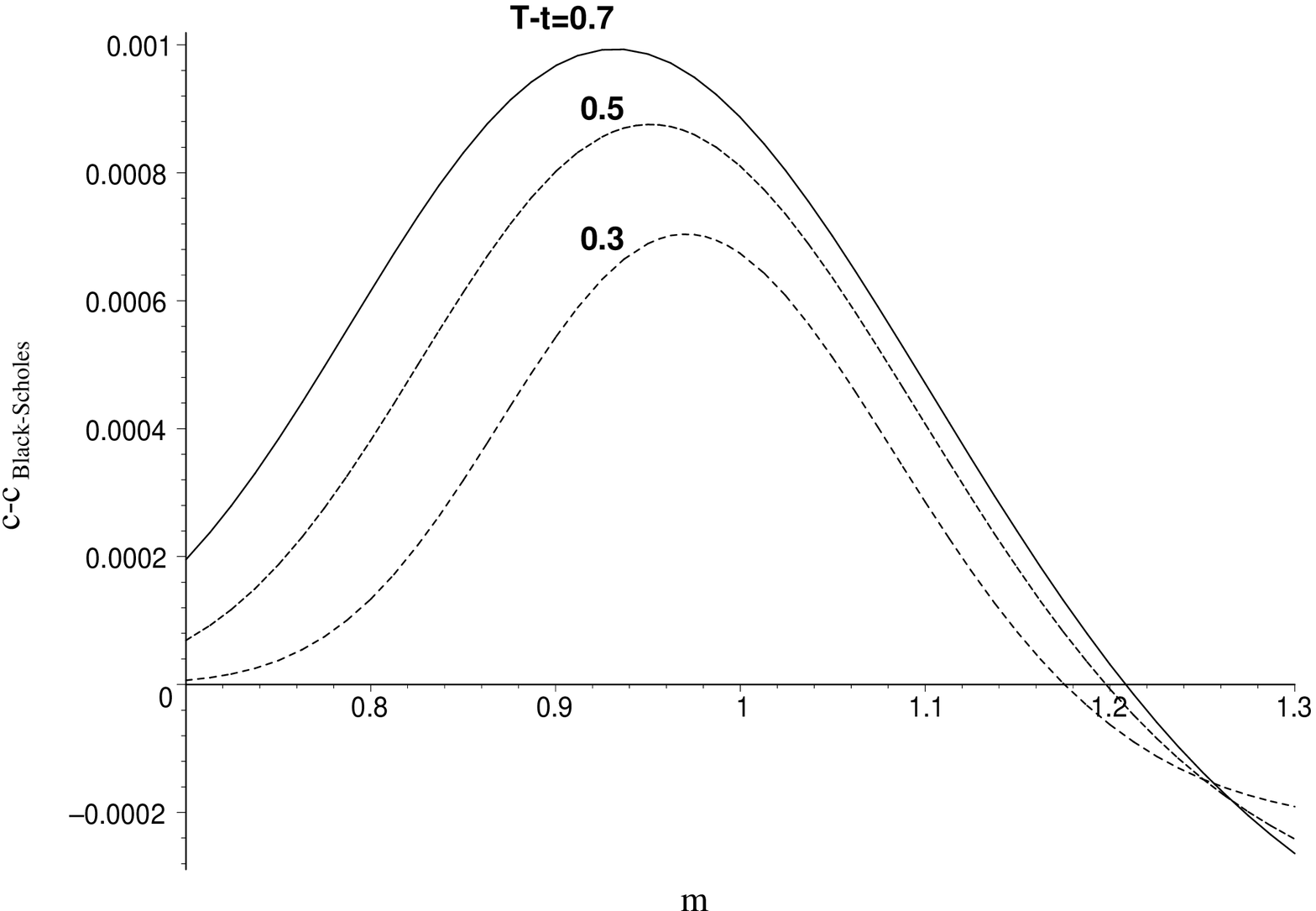,angle=90}
  \end{center}
  \caption{}
  \label{dc1}
\end{figure}

\begin{figure}
  \begin{center}
  \epsfig{file=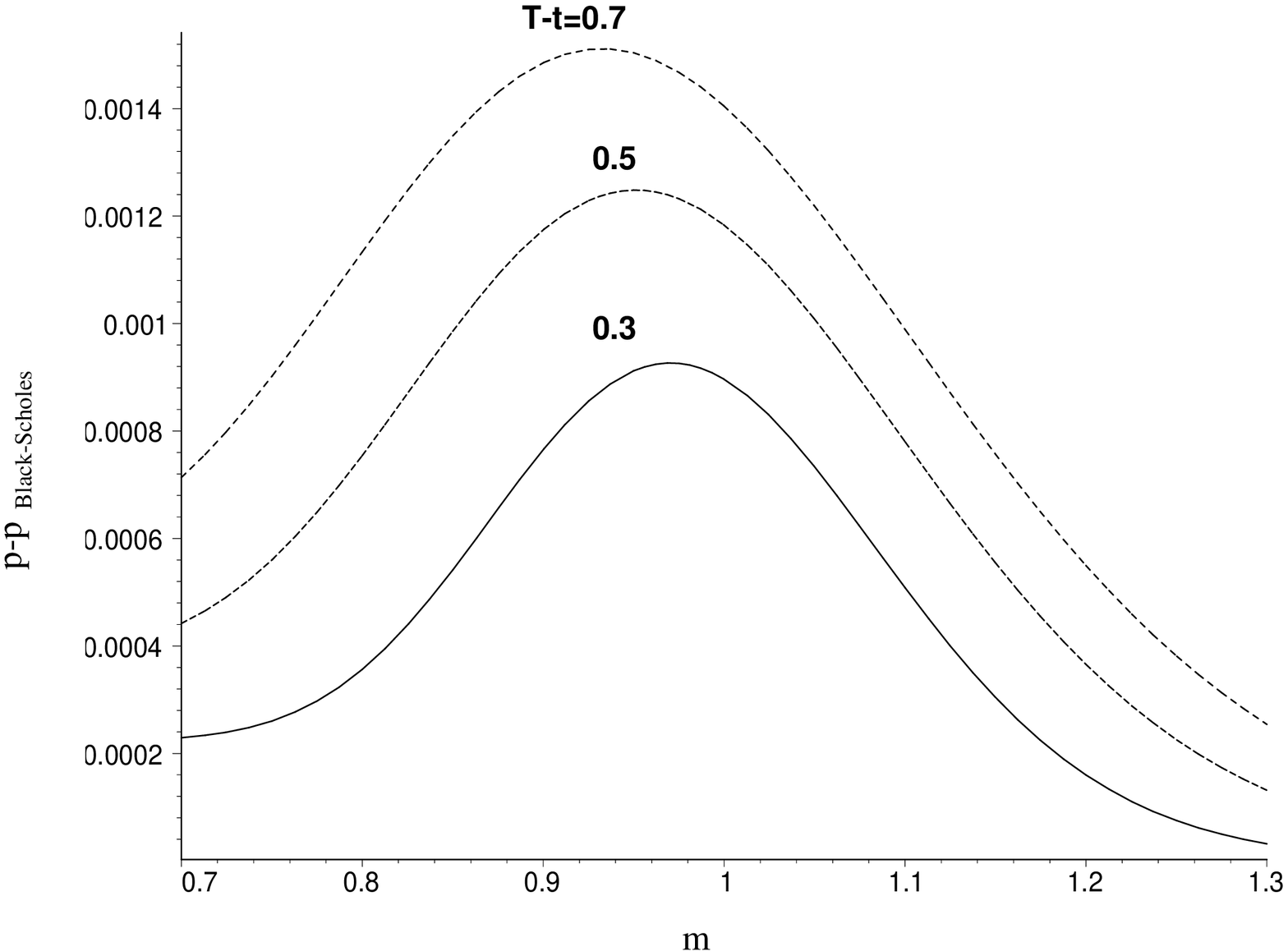,angle=90}  
  \end{center}
  \caption{}
  \label{dp1}
\end{figure}

\newpage
\noindent
FIG. \ref{call1}: The call option price as a function of moneyness
$m=S/K$ (dashed curved line: with virtual arbitrage;
solid line: Black-Scholes formula). The dashed straight line is the
payoff function at maturity. Parameters: 
$\lambda=10$,
$T-t=0.8$, $\Sigma=2$, $\sigma=0.2$, $K=100$,
$r^0=0.08$.\\
\\
FIG. \ref{put1}: The put option price as a function of moneyness
$m=S/K$ (dashed curved line: with virtual arbitrage;
solid line: Black-Scholes formula). The dashed straight line is the
payoff function at maturity. Parameters: see FIG. \ref{call1}.\\
\\
FIG. \ref{dc1}: Difference of the call price to Black-Scholes value (in
absolute value) for various values $T-t$.  Other parameters: $\lambda=100$,
$\Sigma=0.4$, $\sigma=0.2$, $K=100$,
$r^0=0.08$.\\
\\
FIG. \ref{dp1}: Difference of the put price to Black-Scholes value (in
absolute value) for various values $T-t$. Other parameters: see
FIG. \ref{dc1}.\\
\\

\newpage

\newpage

\newpage

\newpage

\end{document}